# Physics courses of the Federal University of Santa Catarina : evasion and gender

# A Física da UFSC em Números: Evasão e Gênero

Débora P. Menezes (1), Karina Buss (1), Caio A. Silvano (1), Beatriz N. D'Ávila (1), Celia Anteneodo (2)

(1) Departamento de Física – CFM – Universidade Federal de Santa Catarina, Florianópolis – SC – CP. 476 – CEP 88.040 – 900 – emails: debora.p.m@ufsc.br, karina.buss@gmail.com, caioasilvano@outlook.com, beatriznattrodt@me.com

(2) Departamento de Física – PUC-Rio, Rio de Janeiro, RJ, CEP 22430-060 – email: celia.fis@puc-rio.br

## Abstract

In this work, we analyse all existing data related to the number of incomers and outcomers (who actually obtain the degree) of the following courses offered at the Federal University of Santa Catarina: physics teaching, bachelor in physics, master of sciences in physics and doctorate in physics, corresponding to the 1998-2017 period, according to their availability.

The data point towards a great male predominance (larger than 76%) and a huge evasion of both sexes (in average, less than 20% of the undergraduate incomers obtain the degree), the evasion being lower in the postgraduate courses and always slightly higher for women.

The average number of incomers and outcomers per year decreases as the students advance from graduate to postgraduate courses, although many students in the postgraduate courses come from other institutions. The proportion of women decreases as the carrier advances.

The results indicate the need of complementary studies that can help the identification of the causes of such a high evasion in order to minimize them.

## Resumo

Neste artigo, são analisados todos os dados existentes relativos aos números de ingressantes e concluintes dos cursos de licenciatura, bacharelado, mestrado e doutorado em física da UFSC, correspondentes ao período de 1988 a 2017, conforme a sua disponibilidade.

Os dados apontam para uma grande predominância masculina (maior que 76%) e deserção de ambos os sexos (em média, menos de 20% dos ingressantes concluem a graduação), sendo a deserção muito menor na pós-graduação e em todos os casos ligeiramente maior para as mulheres.

Os números médios de ingressantes e de concluintes por ano (calculados sobre todo o período considerado) diminuem conforme se avança desde a graduação para a pós-graduação, apesar do fato de que muitos dos alunos ingressantes na pós-graduação vêm de outras instituições. A proporção de mulheres diminui conforme a carreira avança.

Os resultados indicam a necessidade de estudos complementares que identifiquem as causas e as minimizem.

## Introdução

Segundo o Instituto Brasileiro de Geografia e Estatística (IBGE) [1], a composição da população brasileira por sexo é medida pela razão de sexo, que representa o número de pessoas do sexo masculino por cada 100 pessoas do feminino, sendo calculada como o quociente entre o número total de pessoas do sexo masculino dividido pelo número total de pessoas do sexo feminino vezes 100. A razão de sexo obtida em 2014 foi de 93,9, refletindo a composição por sexo de 51,6% de mulheres e 48,4% de homens. Também, segundo o IBGE [1], é maior a taxa de conclusão do ensino médio para mulheres (66,9%) do que para homens (54,9%). No entanto, quando se examinam os percentuais de mulheres que ingressam nas carreiras ligadas às ciências exatas, como no caso da física, e o seu desempenho com o passar do tempo, os números são bem diferentes, como discutiremos neste trabalho.

Um estudo recente realizado pelo Grupo de Trabalho sobre Questões de Gênero (GTG) da Sociedade Brasileira de Física [2] avaliou a intensidade do *efeito tesoura*, tanto nas premiações das Olimpíadas Brasileiras de Física (OBF), quanto nas bolsas de diferentes modalidades do CNPq, desde iniciação científica, oferecidas para estudantes ainda na graduação, até produtividade em pesquisa, ofertadas aos pesquisadores que se destacam entre seus pares. O termo *efeito tesoura* é normalmente utilizado em referência aos gráficos que possuem duas curvas que se cruzam, lembrando uma tesoura aberta. Nas nossas análises, essas curvas estão associadas aos percentuais de homens (que tendem a crescer) e de mulheres (que, complementarmente, tendem a decrescer).

Motivados pelo trabalho supracitado, resolvemos fazer uma análise dos números referentes aos ingressantes e concluintes em todos os cursos de graduação e pós-graduação de física da UFSC, isto é, licenciatura, bacharelado, mestrado e doutorado (separados por sexo) no período de 2000.1 até 2017.1 para a graduação e 1988 a 2017.1 para a pós-graduação, em função dos dados disponíveis.

É importante salientar que a UFSC oferece duas oportunidades de ingresso aos estudantes (uma em cada semestre) e, por isso, os semestres são identificados com índices .1 e .2 depois do ano; por exemplo 2010.1 se refere ao primeiro semestre de 2010 e 2010.2 ao segundo semestre de 2010. Nos dados analisados, em geral, os números relativos aos dois semestres do mesmo ano foram somados, excetuando-se o ano de 2017, para o qual apenas os dados relativos ao primeiro semestre (2017.1) estavam disponíveis quando este artigo foi escrito. Cabe mencionar ainda, que o curso de licenciatura em física é um curso noturno, o curso de bacharelado é um curso diurno e a maioria absoluta dos alunos de mestrado (79,3%) e doutorado (81,6%) recebem bolsas dos órgãos de fomento brasileiros. Também é importante salientar que há sempre uma defasagem entre o número de alunos ingressantes e concluintes decorrente do fato de que os cursos de graduação levam de 4 (bacharelado) a 5 (licenciatura) anos para serem concluídos e os de pós-graduação de 2 (mestrado) a 4 (doutorado) anos. Esse fato não será levado em conta numa primeira discussão sobre a taxa percentual de sucesso nos cursos, definida como o quociente entre os números totais de concluintes e ingressantes, vezes 100, mas será considerado de forma individualizada nas discussões sobre cada curso.

*Licenciatura em Física*

As figuras 1 e 2 representam, respectivamente, os dados de ingresso e de conclusão no curso de licenciatura em Física na UFSC, dentro do período compreendido entre 2000 e 2017.1.

Em cada figura, quando comparados os dados de mulheres (em azul) e homens (em laranja), é visível uma grande disparidade, tanto no caso de ingressantes quanto de concluintes. O efeito tesoura não é quantitativamente drástico, mas está presente, como pode ser inferido comparando os círculos à direita nas figuras 1 e 2 em que se mostra a composição por sexo, e observando que a proporção de mulheres cai em 1,1%.

Os números totais de mulheres e homens ingressantes no curso de licenciatura em física durante todo o período analisado foram respectivamente de 458 e 1488 e os números de concluintes de 77 e 267. Calculamos, então, que as taxas de sucesso são de apenas 17,7 % para mulheres e 19 % para homens, levando-se em conta os números globais de ingressantes, conforme se observa na figura 12. Entretanto, para realizarmos uma estimativa mais rigorosa, retiramos da contagem todos os que ingressaram nos últimos 4 anos porque ainda não poderiam, pelas normas existentes, ter se formado. Foram, então, excluídos 118 mulheres e 330 homens, totalizando 448 estudantes. Encontramos uma taxa de sucesso média de 23%, novamente com a das mulheres (22,7%) ligeiramente abaixo da dos homens (23,1%).

Ainda, comparando as figuras 1 e 2, vemos que em quanto os números de ingressantes por ano oscilam em torno de um valor praticamente constante, os de concluintes apresentam um máximo e uma forte queda depois de 2006.

*Bacharelado em Física*

As figuras 3 e 4 representam os dados, separados por sexo, de ingressantes e concluintes do curso de bacharelado em física na UFSC no período de 2000 até 2017.1. Os números do bacharelado são inferiores aos da licenciatura, porém os valores relativos são próximos, tanto nos percentuais de mulheres quando na taxa de sucesso. Assim como no caso da licenciatura, vemos que a taxa de ingressantes e de concluintes por ano é muito maior entre os homens (em laranja) do que entre as mulheres (em azul). Novamente, comparando os círculos a direita, repara-se que o efeito tesoura, mesmo sendo pequeno, está presente, com a proporção de mulheres diminuindo em 3.3%, diferença ainda maior que no caso da licenciatura.

Enquanto o número de homens que ingressam por ano se mantém praticamente constante dentro de algumas flutuações o número de mulheres que ingressa a cada ano tende a aumentar com o tempo (aproximadamente uma mulher por ano em média). Porém, essa clara tendência praticamente desaparece quando observados os dados de concluintes.

Os números totais de mulheres e homens ingressantes no curso de bacharelado em física de 2000 a 2017.1 foram respectivamente de 310 e 1042 e os números de concluintes de 43 e 176. Comparando os números nos gráficos das figuras 3 e 4, vemos que os de concluintes são notavelmente menores que os de ingressantes, para ambos os sexos e que a evasão é, como no caso da licenciatura, altíssima! Como se pode ver na figura 12, resulta que apenas 13,4% das mulheres e 16,8% dos homens ingressantes concluem o curso. Ao eliminarmos os ingressantes dos últimos 4 anos, 90 mulheres e 218 homens, totalizando 308 estudantes, para uma nova estimativa mais conservadora, o resultado aponta para uma taxa de sucesso de 19,6 % entre as mulheres e de 21,4% para os homens. Todos estes números indicam um insucesso ainda maior no caso do bacharelado em comparação com a licenciatura.

Na figura 5, resumimos os números médios de ingressantes e concluintes, por ano, conjuntamente para os cursos de licenciatura e bacharelado em física da UFSC, calculados sobre todos os anos para os quais existem dados. Qualquer que seja a abordagem adotada, excluindo-se ou não os possíveis formandos, os números apresentados acima para os dois cursos da física indicam a necessidade de fazer algumas reflexões e, necessariamente, estudos mais aprofundados para conhecer os motivos da evasão. A atração que as áreas de ciências exatas exerce nos estudantes do ensino médio é certamente pequena, mas qual a razão que leva um número tão grande de estudantes a desistir dessa formação? Uma tentativa de analisar esses dados será feita nas considerações finais deste artigo e ponderações mais especulativas são abordadas em [3].

*Pós-Graduação (stricto senso) em Física*

Passemos agora a analisar os dados do mestrado e doutorado. As figuras 6 e 7 mostram, respectivamente, os dados dos ingressantes e concluintes no mestrado e as figuras 8 e 9, os ingressantes e concluintes no doutorado em períodos que vão de 1988 a 2017, conforme os dados disponíveis e o período de existência do curso, uma vez que o doutorado foi credenciado pela Capes alguns anos depois do mestrado. Os gráficos apresentados nas figuras 6 e 8 mostram que em vários anos, não houve mulheres ingressantes no PPGFSC.

Cabe ressaltar que uma dissertação de mestrado leva, em média, 24 meses para ser concluída e uma tese de doutorado leva, em média, mais de 48 meses. Os últimos dados do Programa de Pós-Graduação em Física da UFSC (PPGFSC) indicam médias de 23,9 meses para a conclusão do mestrado e de 52,5 meses para a conclusão do doutorado.

Observa-se nos cursos de pós-graduação, novamente, um quadro de evasão ou de insucesso, tanto feminino quanto masculino, porém muito menor que no caso da graduação. As taxas de sucesso (que não levam em conta o tempo necessário para a formação) são de 71,4% e 43,8% para as mulheres no mestrado e doutorado e de 85% e 55,3% para os homens no mestrado e doutorado, conforme mostrado na figura 12. Portanto, o efeito tesoura também se faz presente, sendo as taxas de sucesso das mulheres sistematicamente inferiores às dos homens. Observa-se também, que apesar do longo tempo exigido para titulação no doutorado, os percentuais de insucesso poderiam ser menores. Dados da PPGFSC indicam que o número total de alunos desligados no mestrado foi de 62 e no doutorado de 34. Se descontarmos os alunos que ingressaram no mestrado nos últimos 2 anos e no doutorado nos últimos 4 anos (ainda não tiverem tempo de se graduar), o percentual de evasão médio do mestrado é de (62/(342- 25)= 19,6%) e do doutorado é de (34/(212-49)=20,9%). Portanto, as taxas de sucesso são de 80,4% e 79,1% respectivamente, com percentuais muito próximas àqueles dos bolsistas mencionados no início deste artigo, isto é, 79,3% no mestrado e 81,6% no doutorado.

*Situação Geral – Física UFSC*

Resumindo, nas figuras 10 e 11, mostramos a composição percentual por sexos dos ingressantes e concluintes em todos os cursos de física. Observamos claramente tanto o menor interesse das mulheres pela carreira quanto uma diminuição da proporção de mulheres conforme os estudos progridem, no típico recorte da tesoura.

Além do mais, em todas as categorias é observada uma queda na proporção de mulheres concluintes com relação às ingressantes, pequena mas sistemática: principalmente no doutorado, caindo a proporção de 15,3 a 12,6%.

A figura 12 sintetiza a informação sobre as taxas de sucesso (percentual de graduados, sem o desconto do tempo de titulação) em todos os cursos. O que mais chama a atenção é a evasão generalizada (para ambos os sexos) que os cursos de física sofrem, principalmente os da graduação. De forma sistemática, em todas as categorias, o insucesso é maior no caso das mulheres.

O quadro quantitativo apresentado indica a necessidade de um estudo que permita detectar suas causas para uma possível reversão ou minimização, seja por meio de políticas que ajudem os estudantes a sanar suas dificuldades ao longo dos estudos (várias já existentes, que talvez necessitem de avaliação quantitativa), seja da oferta do curso de licenciatura no período diurno ou através de outra medida que venha a ser vislumbrada.

Finalmente, uma análise temporal sobre o ingresso de mulheres nos cursos de graduação em física, nos remete à figura 13. Vemos claramente uma tendência pequena, mas não desprezível, do aumento da proporção de mulheres entre as ingressantes, da ordem de 0,5% ao ano na licenciatura e 0,8% ao ano no bacharelado. Isso pode ser um indicador positivo de que a proporção de mulheres no doutorado é mais baixa do que no mestrado porque o número de mulheres ingressando na pós-graduação tem aumentado e esse efeito ainda não pode ser observado entre as novas doutorandas. Nesse caso, o que estamos aferindo vai no sentido oposto ao recorte do efeito tesoura e deve resultar num aumento do número de mulheres ingressando na carreira em breve, números que poderão ser confrontados com os existentes no Departamento de Física da UFSC. Neste momento, há 13 mulheres contratadas entre 73 professores [4], contabilizando um percentual de 17,8%. Ao analisarmos o número de docentes credenciados no PPGFSC [5], encontramos 6 mulheres num total de 36 professores, sendo o percentual de contribuição feminina nesse programa de 16,6%. Agora, se consideramos o número de bolsistas de produtividade em pesquisa do CNPq, encontramos apenas 1 mulher entre 15 pesquisadores, num percentual que cai ainda mais, para 6,6%. Mais estudos sobre a evolução temporal desses números serão necessários novamente num futuro não muito distante para que o quadro real fique mais claro.

*Algumas reflexões*

A seguir, levantamos algumas hipóteses, um pouco especulativas, a respeito da situação numérica aqui apresentada. Cabe salientar que entrar nos cursos de graduação em física, tanto licenciatura quanto bacharelado, é facílimo, dada a sua pouca procura com relação à oferta de vagas, mas concluir exige habilidades como raciocínio abstrato bem desenvolvido, vocação para fazer cálculos e muita motivação para estudos avançados. Cabe aqui mencionar que os fatores reais da evasão não são conhecidos e nunca foram mensurados, apesar de sabermos (empiricamente) que a reprovação nas primeiras fases é altíssima. A desistência voluntária em função da procura de outras carreiras, por exemplo, não pode ser descartada.

A possibilidade de uma escolha inconsistente com as habilidades necessárias deve ser investigada e sugere procurar potenciais falhas no material apresentado no ensino médio, o que poderia decorrer da má formação dos professores das escolas, que por sua vez poderia decorrer de falhas estruturais no curso de licenciatura, apontando para a necessidade de uma atenção especial para esse curso.

Os baixos salários dos professores de ensino médio praticados no Brasil, certamente, não servem de estímulo para os alunos que ingressam na licenciatura. Qual, será então, a razão da escolha desse curso e qual a razão da posterior desistência? A percepção não esperada de que o curso demanda uma rotina de estudo muito árdua para pessoas que já trabalham o dia todo (lembrando que o curso é noturno)? Entretanto, observemos que no bacharelado, visando um carreira cientifica, e mesmo sendo um curso diurno, as taxas de insucesso são ainda maiores. Será a causa a falta de cabedal matemático prévio que viabilize o acompanhamento das aulas? Para responder algumas destas perguntas seria interessante ter acesso a dados que permitam identificar as reais causas da evasão.

Com relação à carreira de cientista, as taxas de sucesso são bem maiores que na graduação, em parte devido ao maior compromisso que implica o benefício de uma bolsa de mestrado ou doutorado. No entanto, o insucesso ainda está presente e algumas das perguntas feitas no caso da graduação também se aplicam neste caso. A desistência da carreira científica é um grande problema se pensamos que mesmo que outros setores sejam mais bem remunerados e ofereçam melhores condições de trabalho, os profissionais são sempre formados nas universidades.

Novamente no caso da pós-graduação, seria necessário poder ter acesso a informações mais detalhadas para identificar as causas de fracasso ou desistência e poder intervir adequadamente.

# Referências

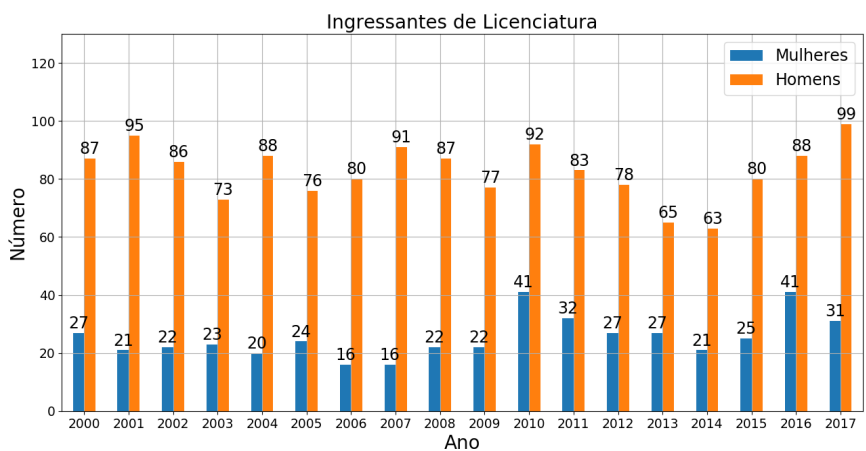
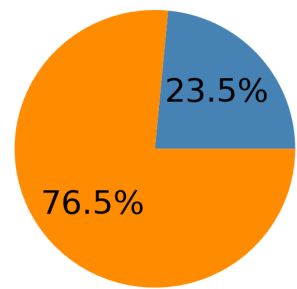

Fig 1 – Número de **ingressantes** (separados por sexo) no curso de **licenciatura em física** da UFSC de 2000.1 a 2017.1. Círculo à direita: porcentagem de ingressantes de cada sexo no mesmo curso e período.

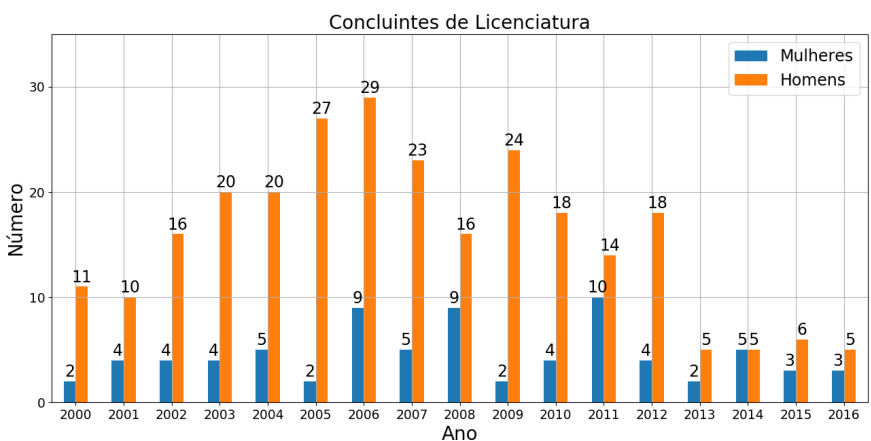
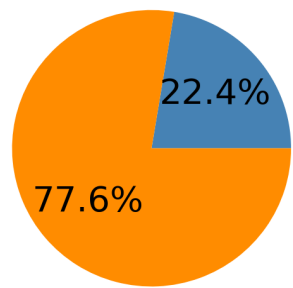

Fig 2 – Número de **concluintes** (separados por sexo) no curso de **licenciatura em física** da UFSC de 2000.1 a 2016.1. Círculo à direita: porcentagem de concluintes de cada sexo no mesmo curso e período.

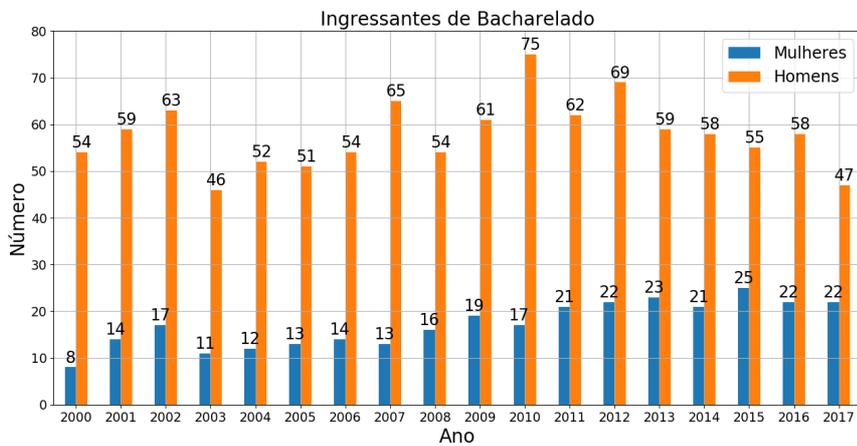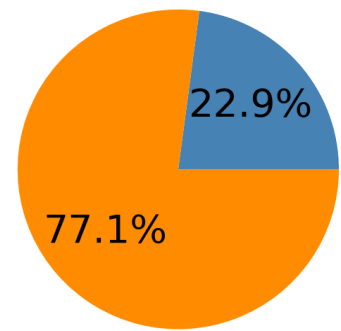

Fig 3 – Número de **ingressantes** (separados por sexo) no curso de **bacharelado em física** da UFSC de 2000.1 a 2017.1. Círculo à direita: porcentagem de ingressantes de cada sexo no mesmo curso e período.

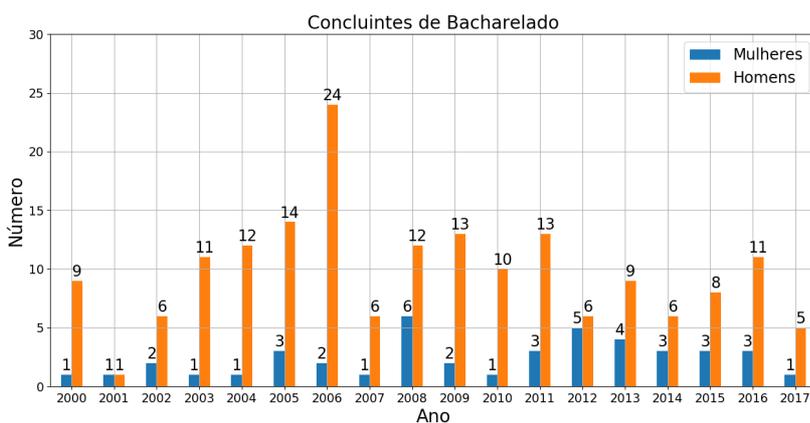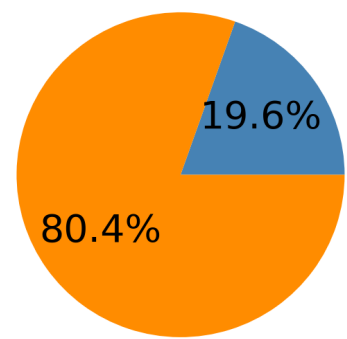

Fig 4 – Número de **concluintes** (separados por sexo) no curso de **bacharelado em física** da UFSC de 2000.1 a 2017.1. Círculo à direita: porcentagem de concluintes de cada sexo no mesmo curso e período.

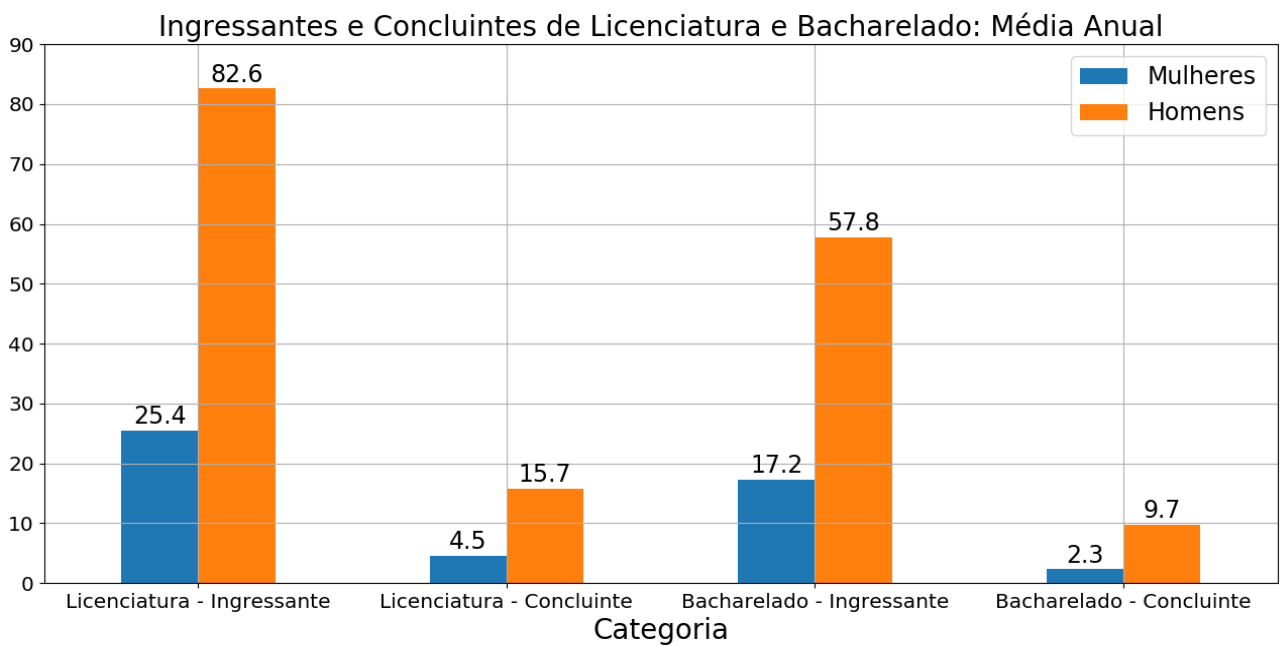
Fig 5 – Número médio (calculado sobre todos os anos disponíveis) de ingressantes e concluintes, de cada sexo, nos cursos de licenciatura e bacharelado em física da UFSC.

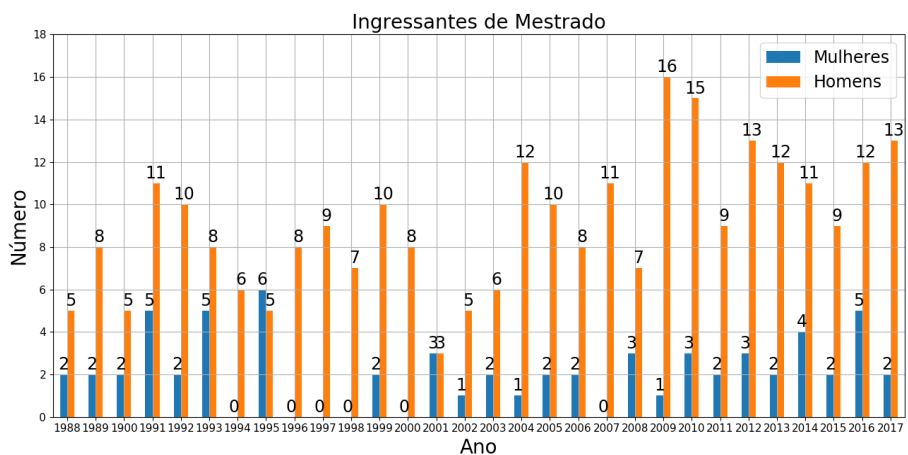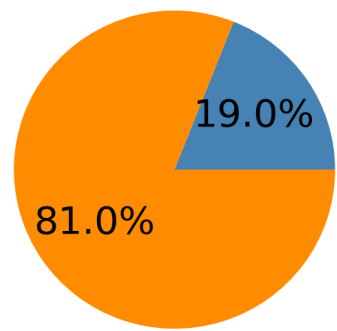

Fig 6 – Número de **ingressantes** (separados por sexo) no **mestrado em física** da UFSC de 1988 a 2017. Círculo à direita: porcentagem de concluintes de cada sexo no mesmo curso e período.

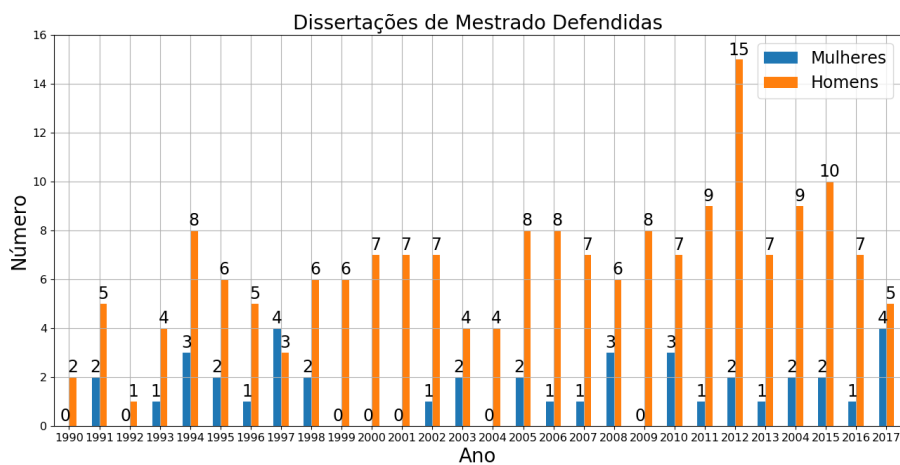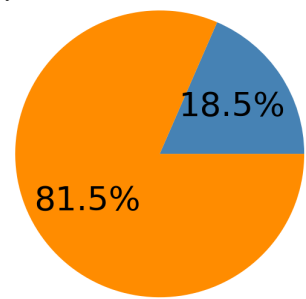

Fig 7 – Número de **dissertações de mestrado em física** defendidas por alunos de cada sexo na UFSC de 1990 a 2017. Círculo à direita: porcentagem de concluintes de cada sexo no mesmo curso e período.

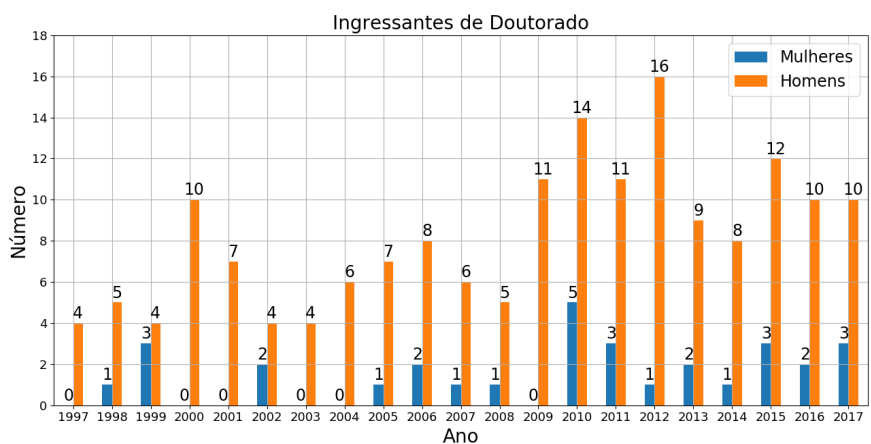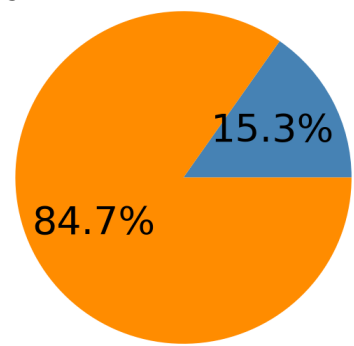

Fig 8 – Número de **ingressantes** (separados por sexo) no **doutorado em física** da UFSC de 1997 a 2017. Círculo à direita: porcentagem de concluintes de cada sexo no mesmo curso e período

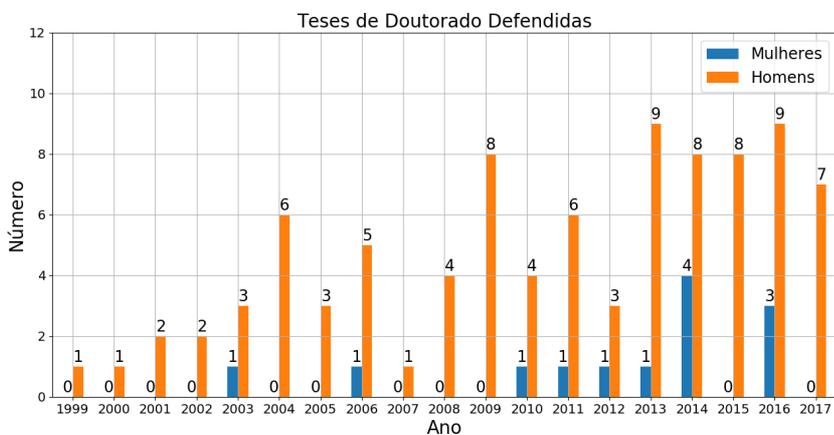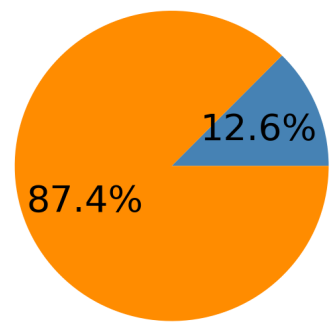

Fig 9 – Número de **teses de doutorado em física defendidas** na UFSC entre 1999 a 2017, por alunos de cada sexo. Círculo à direita: porcentagem de concluintes de cada sexo no mesmo curso e período.

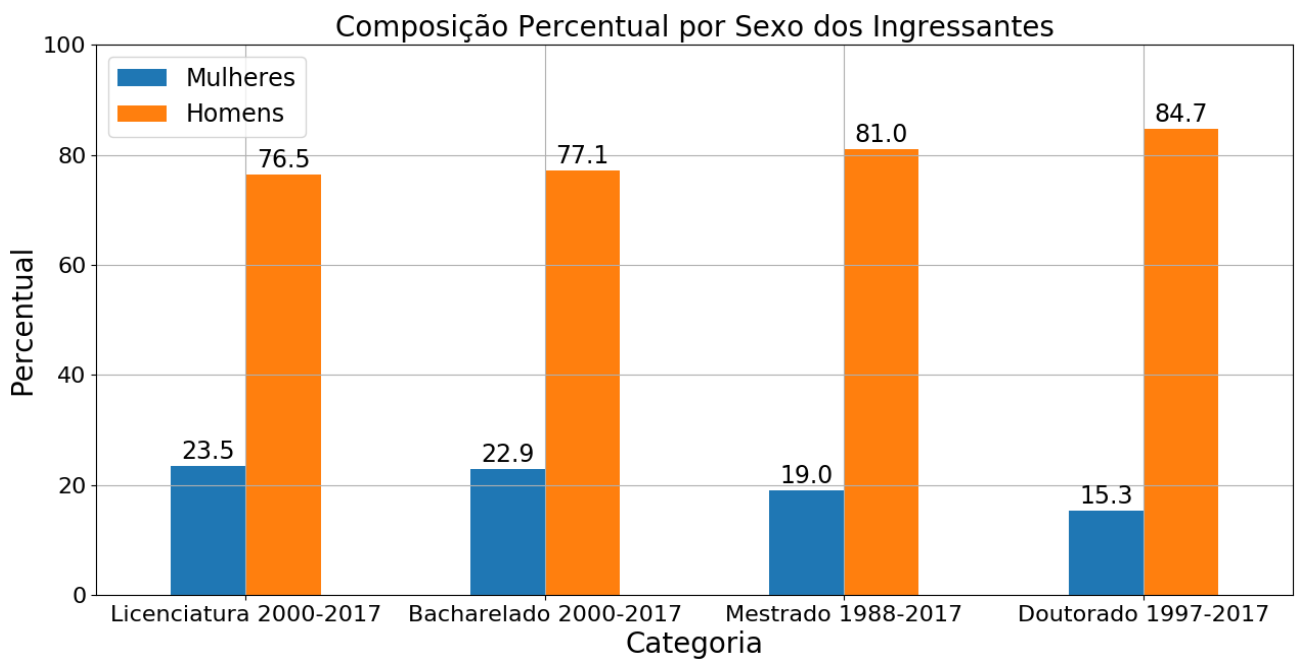

Fig 10 – Percentual médio de ingressantes de cada sexo nos diversos cursos de física da UFSC.

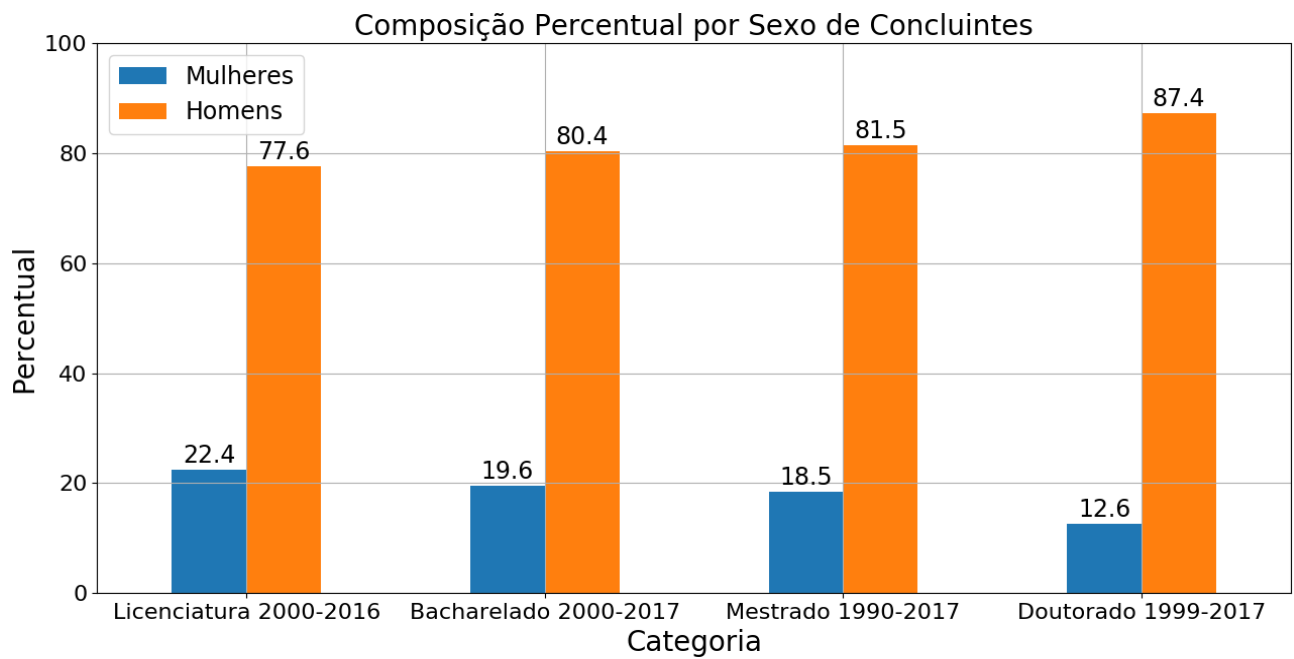

Fig 11 – Percentual médio de concluintes de cada sexo nos diversos cursos de física da UFSC.

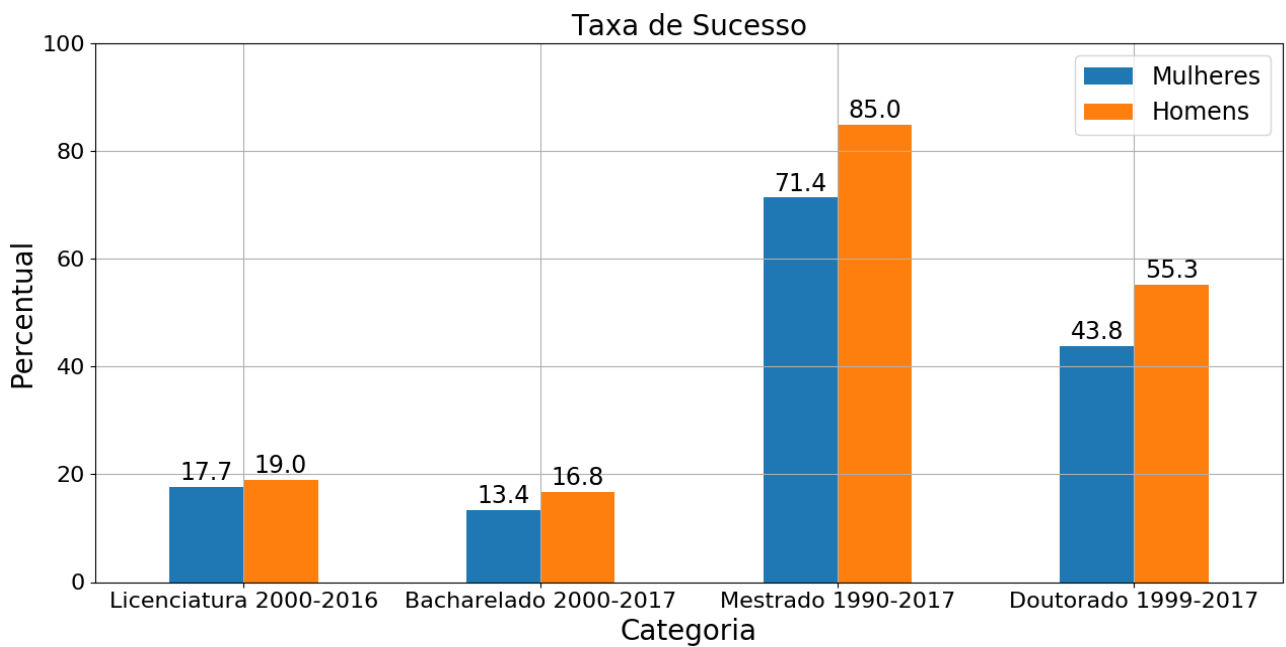

Fig 12 – Taxa de sucesso nos diversos cursos de física da UFSC para estudantes de ambos os sexos.

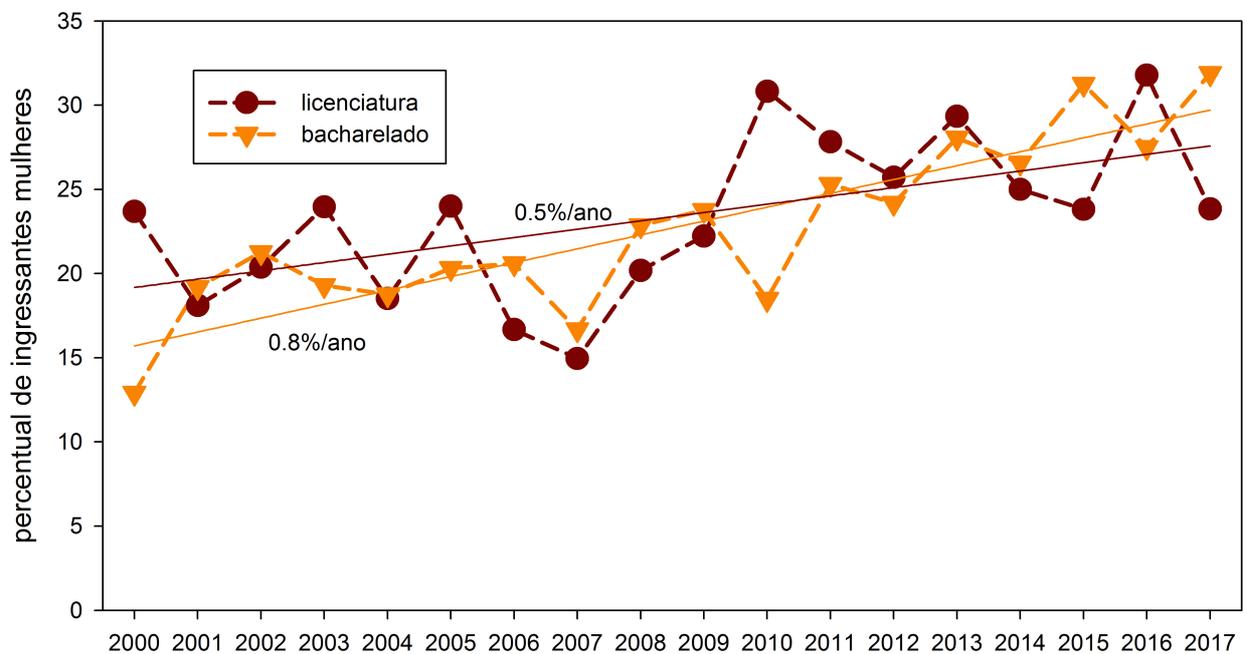

Fig 13 - Percentual de ingressantes mulheres nos cursos de bacharelado e licenciatura de física da UFSC. Análise estatística.